\documentclass[utf8]{frontiersSCNS} % for Science, Engineering and Humanities and Social Sciences 
\pdfoutput=1
\usepackage{url,hyperref,lineno,microtype,subcaption}
\usepackage[onehalfspacing]{setspace}

\def\keyFont{\fontsize{8}{11}\helveticabold }
\def\firstAuthorLast{Aryaman {et~al.}} %use et al only if is more than 1 author
\def\Authors{Juvid Aryaman\,$^{1,2,3}$, Iain G.\ Johnston\,$^{4,5}$ and Nick S. Jones\,$^{1,5,*}$}
\def\Address{$^{1}$Department of Mathematics, Imperial College London, London, United Kingdom\\
$^{2}$ Department of Clinical Neurosciences, University of Cambridge, Cambridge, United Kingdom\\
$^{3}$ MRC Mitochondrial Biology Unit, University of Cambridge, Cambridge, United Kingdom\\
$^{4}$School of Biosciences, University of Birmingham, Birmingham, United Kingdom\\
$^{5}$EPSRC Centre for the Mathematics of Precision Healthcare, Imperial College London, London, United Kingdom }

\def\corrAuthor{Nick Jones}

\def\corrEmail{nick.jones@imperial.ac.uk}

\newcommand{\dpsi}{\Delta \Psi}

\newcommand{\ca}{Ca\textsuperscript{2+}}

\begin{document}
\onecolumn
\firstpage{1}

\title[Mitochondrial heterogeneity]{Mitochondrial heterogeneity} 

\author[\firstAuthorLast ]{\Authors} %This field will be automatically populated
\address{} %This field will be automatically populated
\correspondance{} %This field will be automatically populated

\extraAuth{}

\maketitle

\begin{abstract}

Cell-to-cell heterogeneity drives a range of (patho)physiologically important phenomena, such as cell fate and chemotherapeutic resistance. The role of metabolism, and particularly mitochondria, is increasingly being recognised as an important explanatory factor in cell-to-cell heterogeneity. Most eukaryotic cells possess a population of mitochondria, in the sense that mitochondrial DNA (mtDNA) is held in multiple copies per cell, where the sequence of each molecule can vary. Hence intra-cellular mitochondrial heterogeneity is possible, which can induce inter-cellular mitochondrial heterogeneity, and may drive aspects of cellular noise. In this review, we discuss sources of mitochondrial heterogeneity (variations between mitochondria in the same cell, and mitochondrial variations between supposedly identical cells) from both genetic and non-genetic perspectives, and mitochondrial genotype-phenotype links. We discuss the apparent homeostasis of mtDNA copy number, the observation of pervasive intra-cellular mtDNA mutation (we term `microheteroplasmy'), and developments in the understanding of inter-cellular mtDNA mutation (`macroheteroplasmy'). We point to the relationship between mitochondrial supercomplexes, cristal structure, pH, and cardiolipin as a potential amplifier of the mitochondrial genotype-phenotype link. We also discuss mitochondrial membrane potential and networks as sources of mitochondrial heterogeneity, and their influence upon the mitochondrial genome. Finally, we revisit the idea of mitochondrial complementation as a means of dampening mitochondrial genotype-phenotype links in light of recent experimental developments. The diverse sources of mitochondrial heterogeneity, as well as their increasingly recognised role in contributing to cellular heterogeneity, highlights the need for future single-cell mitochondrial measurements in the context of cellular noise studies.

\tiny
 \keyFont{ \section{Keywords:} Mitochondria, Microheteroplasmy, Macroheteroplasmy, Genotype-phenotype Links, Threshold Effect, Complementation, Cellular Noise, Cell-to-cell Heterogeneity, Heteroplasmy Variance} 
\end{abstract}

\section*{Introduction}

Cellular heterogeneity plays central functional roles in a variety of biomedically important phenomena, such as development \citep{Vassar93,Chang08}, virus infection \citep{Snijder09}, chemotherapeutic resistance \citep{Spencer09,Marquez18}, and gene expression in ageing \citep{Bahar06}. Inter-cellular heterogeneity may arise from the intrinsically stochastic nature of cellular processes including gene transcription \citep{Swain02,Elowitz02}, but also from `extrinsic' sources such as the cell cycle \citep{Newman06} and partitioning noise \citep{Huh11b,Huh11,Johnston15b}. 

We define mitochondrial heterogeneity as (1) the variation of a mitochondrial feature within the mitochondrial population of a single cell, and (2) the variation of a mitochondrial feature, potentially aggregated at the per-cell level, between supposedly identical cells. Mitochondrial heterogeneity has been found to be an important correlate of extrinsic cellular noise. \cite{Das2010} and \cite{Johnston12} found that mitochondrial mass, scaled by mitochondrial membrane potential, correlates strongly with global transcription rate at the single-cell level (also accounting for $\sim 50\%$ of the heterogeneity observed in protein levels \citep{Guantes15}). Since global transcription rate has diverse implications for cellular function \citep{Raj08}, this provides compelling evidence for the importance of mitochondrial heterogeneity as a contributor to cellular noise. Indeed, mathematical modelling predicted that mitochondrial functionality influences stem cell differentiation \citep{Johnston12}, which has received experimental support \citep{Gaal14,Sukumar16,Anso17}. Furthermore, mitochondrial heterogeneity has been correlated with cell-to-cell heterogeneity in chemotherapeutic resistance in mammalian cells \citep{Mizutani09,Yao13,Marquez18}, as well as proliferation rate and drug resistance in yeast \citep{Dhar18}. While many existing studies on mitochondrial heterogeneity are correlative, rather than showing causal mechanisms, the importance of metabolic noise is becoming more widely recognised, including in prokaryotic systems \citep{Kiviet14,Takhaveev18}.

\begin{figure}
\begin{center}
\includegraphics[width=0.85\columnwidth]{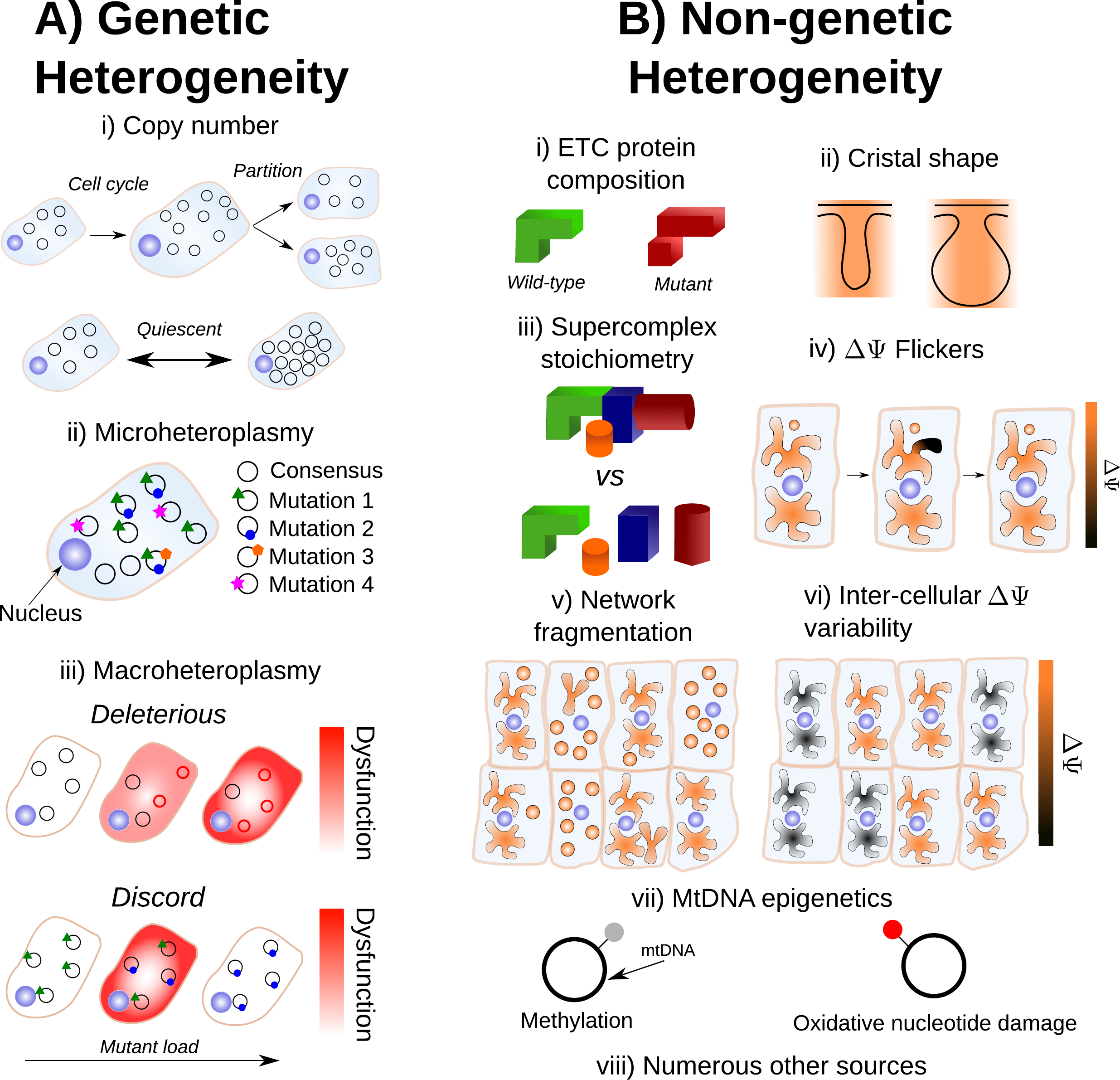}  
\end{center}
\caption{ Sources of mitochondrial heterogeneity from genetic and non-genetic mitochondrial sources. \textbf{A}.~Potential sources of mitochondrial genetic heterogeneity. i) A proliferative cell which doubles in volume is expected to replicate its mtDNA complement by approximately $\times 2$ to avoid dilution, which may confound measurements of copy number heterogeneity; noise at partitioning is thought to be somewhat suppressed. The extent of copy number heterogeneity in quiescent cells, and its consequences, are not fully understood. ii) A cloud of mutations around the consensus sequence is expected given a finite mtDNA mutation rate; we term this `microheteroplasmy'. The actual mutant proportion for most mutations is expected to be very small in reality. iii) In the more canonical case, `macroheteroplasmy', a deleterious sequence (such as a deleterious single nucleotide polymorphism or deletion, red circles) induces a pathological phenotype above a threshold heteroplasmy. By contrast, intermediate heteroplasmy has been observed to induce large fitness disadvantages but homoplasmy does not, indicating discord between the two alleles. \textbf{B}.~Potential sources of mitochondrial non-genetic heterogeneity. i) MtDNA sequence variation can induce variation in the structure of corresponding proteins. ii) Cristal structure is variable and physiologically regulated to control respiratory output. iii) A landscape of supercomplex stoichiometries exists. iv) Spatially-restricted, transient, depolarisation/repolarisation cycles of the mitochondrial network are observed. v) The extent of mitochondrial fragmentation influences heteroplasmy dynamics (see Eq.\eqref{eq:my_eq}). vi) Intercellular heterogeneity in mitochondrial membrane potential has been shown to be an important source of cellular noise. vii)~Potential sources of mitochondrial epigenetic heterogeneity. MtDNA is observed to undergo epigenetic modification by methylation, although the physiological significance of this is uncertain. Also, oxidative damage to mtDNA nucleotides may cause transcriptional errors. viii) There are numerous other sources of non-genetic heterogeneity: from distinct spatial niches to ER-mitochondrial interactions. For further discussion, see Main Text.
}
\label{Fig:cartoon}
\end{figure}

In this review, we discuss sources of mitochondrial heterogeneity. Since individual cells consist of a population of mitochondria, intra-cellular heterogeneity may exist and may give rise to inter-cellular heterogeneity. Mitochondria also display a rich physiology, apart from genetic aspects. We therefore partition our discussion into genetic and non-genetic sources of mitochondrial heterogeneity and indicate their inter-dependence through genotype-phenotype links (Figure~\ref{Fig:cartoon}). Our discussion is biomedically focussed, so we will emphasise findings in humans and model organisms used to study human disease such as mouse, fly, worm and yeast (the plant kingdom has several unique aspects of mitochondrial heterogeneity, recently reviewed by \cite{Johnston18}). We provide a necessarily non-exhaustive discussion of recent developments in these topics, and point out potential areas for future development.

\section*{Genetic Sources of Mitochondrial Heterogeneity}

Genetic and non-genetic sources of mitochondrial heterogeneity are not independent. Variation in the amount and sequence of mtDNA affects the number and sequence of corresponding transcripts, eventually affecting respiratory output (although these relationships may be non-linear \citep{Rossignol03,Rocher08,Picard14,Aryaman17b}). Conversely, the non-genetic state of the cell may affect the mitochondrial genetic state, for instance AMP/ATP ratios may alter mitochondrial biogenesis and autophagy (mitophagy) rates \citep{Palikaras14}. Variation in mtDNA turnover rates have the potential to affect not only mtDNA copy number but also heteroplasmy (defined as the fraction of a particular variant allele of mtDNA per cell) through neutral genetic drift \citep{Birky83,Chinnery99,Capps03,Wonnapinij08,Johnston16}, and selective effects \citep{Larsson90,Hart96,Bua06,Ye14,Li15,Morris17,Floros18}, as we discuss below.

\subsection*{Inter-cellular MtDNA Copy Number Appears to be under Homeostatic Control}

\textit{MtDNA copy number appears to be under homeostatic control}. A potential source of mitochondrial genotypic heterogeneity is simply the quantity of mtDNA per cell. A proliferating cell is expected to increase its complement of mtDNAs by a factor of 2 over the cell cycle, and noisily partition those molecules upon division \citep{Johnston15b} (Figure~\ref{Fig:cartoon}Ai). However, a number of studies suggest that cytoplasmic mtDNA density is tightly controlled within a variety of replicating cell types (although mtDNA copy number, and density, varies radically between different cell types \citep{Wilson01}). Work in HeLa cells \citep{Posakony1977} and budding yeast \citep{Rafelski12} have indicated that mitochondrial volume increases approximately in proportion to cytoplasmic volume, although we note that a recent \textit{in vivo} study involving hypertrophic mouse hepatocytes suggested that mtDNA density may reduce with cell size \citep{Miettinen14}. Further single-cell studies are required to validate this observation. Studies in proliferative human cell lines \citep{Iborra04,Tauber13}, budding yeast \citep{Osman15} and fission yeast \citep{Jajoo16} have shown that the distribution of inter-nucleoid spacings is significantly perturbed from random, suggesting that mtDNA density is controlled in proliferating cells. Indeed, mathematical modelling using a constant mitochondrial density \citep{Johnston12} was able to explain a range of single-cell data for replicating cells \citep{Das2010}.

\textit{Interpretation of apparent mtDNA copy number homeostasis}. The conservation of mitochondrial density is somewhat surprising, given that mitochondrial density is a potential axis for cells to control power production in response to differing demands, especially in the context of differing cell volume. Smaller cells have a larger surface area to volume ratio, therefore power demand is not expected to scale linearly with cell volume. Mathematical modelling has suggested that cells may instead modulate their mitochondrial membrane potential, rather than their mtDNA density, to satisfy cellular demands in mammalian cells \citep{Miettinen16,Aryaman17}, perhaps affording the cell more control since membrane potential may change on a faster timescale than mtDNA biogenesis. The extent to which mtDNA density homeostasis holds in the absence of cell volume variation driven by the cell cycle, i.e.\ quiescent cells, has yet to be carefully explored (Figure~\ref{Fig:cartoon}Ai) despite its relevance for mosaic dysfunction in ageing post-mitotic tissues \citep{Kauppila17}.

\textit{Pathological consequences of loss of mtDNA copy number homeostasis}. In humans, a variety of nuclear mutations which induce defects in mtDNA maintenance cause mitochondrial depletion syndromes; these are severe disorders and clinically diverse in their physiological impact \citep{El03}. Conversely, it has been shown that increasing mtDNA copy number can rescue male infertility in mice engineered to accumulate mtDNA mutations, despite unaltered heteroplasmy \citep{Trifunovic04,Jiang17}. It has been hypothesised that failure to maintain homeostasis in the density of functional mtDNAs may underlie the pathology of one of the most common mtDNA mutations associated with mitochondrial disease (3243A$>$G tRNA mutation) \citep{Aryaman17b}. A mathematical model of human cybrid cells with the 3243A$>$G mutation was consistent with a range of omics data \citep{Picard14} by assuming that cells attempt to maintain mtDNA density homeostasis through cytoplasmic volume reduction, until a minimum cell volume is reached where cells undergo a switch in their metabolic response \citep{Aryaman17b}. Indeed, assuming constant mitochondrial functionality, the study of \cite{Johnston12} predicts that a reduction in mtDNA density results in lowered ATP concentrations, which results in lowered transcription rate \citep{Das2010}. These studies highlight the potential pathophysiological relevance of maintaining mtDNA density homeostasis.

\subsection*{Intra-cellular Mutations in Mitochondrial DNA are a Source of Genotypic Heterogeneity}

\textit{MtDNA mutation as a source of heterogeneity}. Mitochondrial DNA is replicated and degraded, even in non-proliferating tissues, which generates opportunities for mtDNA mutations to arise and proliferate. Studies of mtDNA mutation spectra in humans have suggested that point mutations predominantly arise from replication errors \citep{Williams13,Kennedy13,Stewart14} as opposed to oxidative damage \citep{Kauppila15,Kauppila18}, as is also the case for the ``common'' 4997~bp deletion \citep{Phillips17}.  

\textit{Intra-cellular mtDNA mutation as a source of heterogeneity}. Finite mutation rates during replication of mtDNA are expected to give rise to a set of closely-related sequences which do not all necessarily maximise fitness \citep{Eigen77,Nowak06}. Therefore, at the intra-cellular level, we expect to observe mtDNA sequence diversity (see e.g.~\citep{Jayaprakash15}). Recent experimental work in primary cultures of mouse neurons and astrocytes has shown this to be the case \citep{Morris17}. The authors found 3.9$\pm$5.7 single nucleotide variations (SNVs) ($\pm$ standard deviation) per mitochondrion \citep{Morris17}, with a mitochondrion expected to contain around 5 molecules of mtDNA \citep{Satoh91}. The authors of this study found that the distribution of allele frequencies was skewed towards 0\% heteroplasmy \citep{Morris17}, suggesting the existence of negative selection acting at the intra-cellular level \citep{Birky83}. However, several moderate/high impact mutations with $>$90\% heteroplasmy were also discovered \citep{Morris17}, suggesting that certain mutations are able to evade intra-cellular selection and reach high levels of heteroplasmy. Further studies are required to determine whether such high-heteroplasmy variants become established through neutral drift, or whether a positive selection mechanism exists; for example, it is possible that heterogeneity in the intra-cellular environment establishes intra-cellular niches which favour different mtDNA sequences. We use the term `microheteroplasmy' to denote abundant ultra-low heteroplasmic mutations within single cells (Figure~\ref{Fig:cartoon}Aii).

\textit{Pathological consequences of microheteroplasmy}. Whilst the pathological implications of  microheteroplasmy remain unclear, comparisons could be drawn to  experiments where a naturally occurring, but foreign, mtDNA haplotype is introduced into a cell with which it has not co-evolved. Increasing genetic distance amongst haplotypes has been shown to induce tissue-specific selective pressures in mice \citep{Burgstaller14}, and heteroplasmy between otherwise healthy haplotypes have been shown to induce fitness disadvantages \citep{Acton07,Sharpley12}, as corroborated by bottom-up mathematical modelling \citep{Hoitzing17}. Recent work has shown that such fitness disadvantages may be mediated through oxidative damage \citep{Bagwan18} (Figure~\ref{Fig:cartoon}Aiii). 

\subsection*{Inter-cellular Mutations in Mitochondrial DNA are a Source of Genotypic Heterogeneity}

\begin{figure}
\begin{center}
\includegraphics[width=0.85\columnwidth]{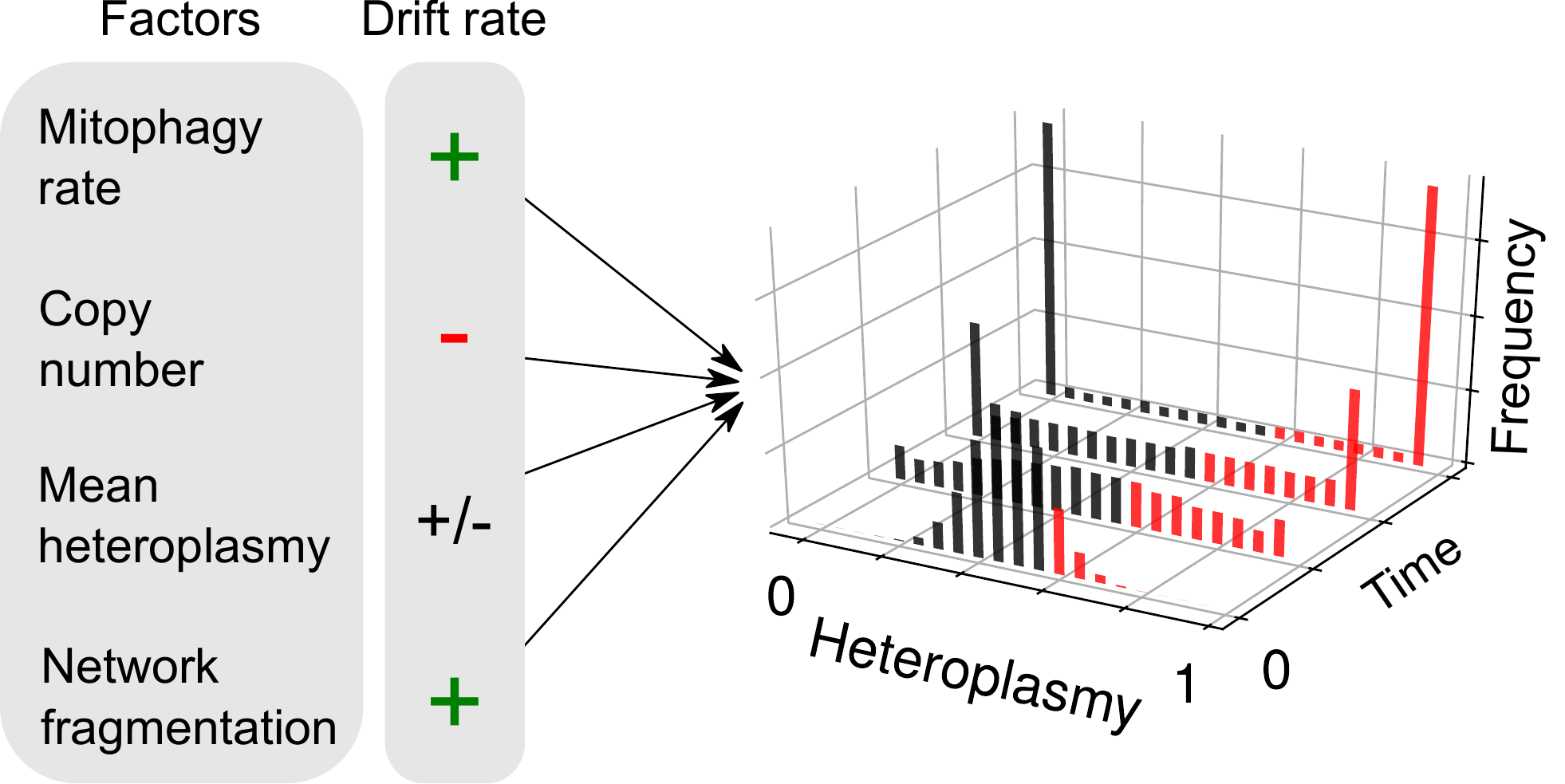}  
\end{center}
\caption{ Factors influencing neutral genetic drift of mtDNA. Heteroplasmy ($h$, the fraction of a particular variant allele of mtDNA per cell) is not generally constant between cells: it is a random variable and yields a distribution of cellular states. If the consensus and variant alleles experience the same instantaneous birth and death rates per cell, then the heteroplasmy distribution is subject to `neutral drift'. Neutral drift is characterised by the increase in variance of the heteroplasmy distribution with time. It is thought that when cells exceed a particular threshold heteroplasmy, a pathological phenotype may be expressed (red bars). Therefore, the number of pathological cells may increase with heteroplasmy variance. Mathematical modelling has shown that the rate of increase of heteroplasmy variance ($\mathbb{V}(h)$) increases with mitophagy rate, as higher turnover provides more opportunities for replication of either allele and cause a change in $h$. $\mathbb{V}(h)$ decreases with copy number, since large populations are more robust to fluctuations. $\mathbb{V}(h)$ is maximal when mean heteroplasmy is 0.5, and diminishes as one allele dominates over the other. It has recently been shown that mitochondrial network fragmentation can rescale the turnover rate: larger fragmentation results in $\mathbb{V}(h)$ increasing faster with time -- as more mitochondria are susceptible to mitophagy -- independently of the absolute fusion-fission rates, see Eq.\eqref{eq:my_eq}.
}
\label{Fig:dists_cartoon}
\end{figure}

\textit{Cell-to-cell heterogeneity in heteroplasmy increases linearly with time}. Another consequence of the finite mutation rate of mtDNA, and its stochastic turnover, is that mutations will occasionally drift to very high level heteroplasmies simply by genetic drift \citep{Birky83,Ewens04}, i.e. `macroheteroplasmy' (Figure~\ref{Fig:cartoon}Aiii). Cell-to-cell heterogeneity in heteroplasmy (heteroplasmy variance) therefore widens through time. Certain deleterious mutations may therefore drift to high levels of heteroplasmy and cause pathology \citep{Rossignol03} (see Figure~\ref{Fig:dists_cartoon}). Theoretical work predicts that heteroplasmy variance increases approximately linearly, and increases with the rate of mitophagic turnover of mtDNAs \citep{Poovathingal09,Johnston16}, but decreases with total mtDNA copy number with a non-linear dependence upon mean heteroplasmy \citep{Johnston16}. A linear increase in heteroplasmy variance with time  has recently been observed in mouse oocytes, and over pup lifetimes \citep{Burgstaller18}. Mathematical modelling has suggested that increasing heteroplasmy variance might increase the energetic cost of maintaining a tissue \citep{Hoitzing17}.

\textit{Heteroplasmy variance and selective thresholds may counter the progressive increase in mutant load during development}. Increases in heteroplasmy variance can also purge mitochondrial mutations when combined with selection, as observed between generations in the human mitochondrial bottleneck \citep{Johnston15,Burr18,Floros18}. During development, through a reduction of cellular mtDNA copy number, heteroplasmy variance is increased. When combined with a selective apoptotic threshold, whereby cells above a heteroplasmic threshold undergo cell death, mean heteroplasmy can be reduced. The mechanisms and timings of the mitochondrial bottleneck are debated \citep{Jenuth96,Cao07,Cree08b,Wai08}; however, stochastic modelling has shown that several proposals are compatible with the induction of heteroplasmy variance through a combination of random mtDNA partitioning at division and passive turnover of mtDNA \citep{Johnston15,Johnston16,Hoitzing17c}.

\textit{Clonal expansions of mitochondrial mutations with age have pathological consequences}.  Mitochondrial mutations which survive the bottleneck can result in severe congenital diseases \citep{Schon12}, and accelerate ageing in the case of severe mutant loads in mice \citep{Ross13}. Under physiological conditions, mtDNA mutations are also observed to accumulate with age across tissues in humans \citep{Khrapko99,Taylor03,Kraytsberg06,Bender06,Li15,Kauppila17}. Certain cases of mutant accumulation (or ``clonal expansion'') may be explained by neutral drift, for instance partitioning noise in highly proliferative colonic crypt cells can result in highly mutated crypts \citep{Taylor03}. However, positive selection through replicative advantage \citep{Samuels13} and tissue-specific niches \citep{Avital12,Li15,Ahier18} can also influence heteroplasmy dynamics.  The extent to which mtDNA mutations present at birth, versus somatic mutation post-development, contribute to healthy ageing remains open. Bulk heteroplasmy measurements in monozygotic twins have shown that inherited mutations can contribute mitochondrial mutations, including heteroplasmic mutations with a low mutant load \citep{Avital12}. The vast amount of DNA replication during development itself is also likely to seed and expand a number of potentially pathological mtDNA mutations \citep{Hahn18}, as is thought to be the case in the nuclear genome \citep{Keogh18b}. Mitochondrial mutations which arise and focally expand post-development in skeletal muscle contribute to pathological age-related loss in muscle mass \citep{Khrapko99,Aiken02,Bua06,Vincent18}. In the case of human muscle fibres of healthy individuals, the number of fibres exhibiting electron transport chain (ETC) abnormalities increases from approximately 6\% at age 49 to 31\% at age 92 \citep{Bua06}. Serial sections through laser-captured single skeletal muscle fibres showed that heteroplasmy in mtDNA deletions may exceed 90\% \citep{Bua06}. Deletion mutations have been associated with local fibre atrophy and breakage in mice \citep{Wanagat01,Khrapko09} and rhesus monkeys \citep{Aiken02} (although the causative role of mtDNA mutations in ageing for shorter-lived animals is contested \citep{Vermulst07,Lakshmanan18,Kauppila18b}). It is noteworthy that mitochondrial dysfunction in a particular tissue is able to induce stress responses in distal tissues through the mitochondrial unfolded protein response \citep{Zhao02,Durieux11,Zhang18} and other hormonal signalling pathways \citep{Tyynismaa10,Khan17}, suggesting potential organism-wide consequences of focal mitochondrial mutations. The development of novel, quantitative, dyes for electron transport chain (ETC) deficiency may allow more refined measurements of the causes of deletion proliferation in somatic tissues in the future \citep{Simard18,Vincent18b}.

\textit{Inter-cellular mitochondrial exchange as a physiological means to slow heteroplasmy variance}. Mitochondria have been observed to be transferred between cells \citep{Spees06,Torralba16}, as has the intercellular transfer of other organelles \citep{Rustom04}. Several mechanisms by which mitochondrial transfer is achieved are known, including tunnelling nanotubes \citep{Koyanagi05} and microvesicles \citep{Phinney15}, amongst others \citep{Torralba16}. Intercellular mitochondrial exchange has largely been understood through in vitro experiments \citep{Spees06,Jayaprakash15,Berridge16}; however, recent developments have shown the effect to also be relevant \textit{in vivo} e.g.\  \citep{Islam12,Ahmad14,Berridge16}. For instance, \cite{Islam12} have shown that mouse mesenchymal stem cells (MSCs) are able to release microvesicles containing mitochondria, and protect against sepsis-induced acute lung injury in mice. Recent evidence from xenograft mouse models have also shown that mtDNA can be transferred between cells via exosomes, mediating an escape from dormancy in therapy-resistant breast cancer cells \citep{Sansone17}. Furthermore, astrocytes have been found to donate mitochondria to neurons after stroke \citep{Hayakawa16}, although interpretations of these data exist which do not require mitochondrial exchange \citep{Berridge16b}. In terms of mitochondrial genetic dynamics, intercellular exchange of mtDNA appears to slow down heteroplasmic drift \citep{Jayaprakash15} and thereby reduce cell-to-cell mitochondrial genetic heterogeneity. It is possible that inter-cellular exchange of mitochondria, and mitochondrial DNA, is an evolved mechanism to ameliorate heteroplasmy variance and therefore the build-up of cells with pathological levels of heteroplasmy. Although the existence of inter-cellular mitochondrial exchange has now been established, the extent and dynamics of mtDNA transfer remain incompletely understood.

\textit{Gene therapy as a means to therapeutically control heteroplasmy distributions}. Gene-editing technologies targeted at the mitochondrial genome are under development \citep{Bacman13,Gammage14,Reddy15,Pereira18,Bacman18,Gammage18}. Such technologies sequence-specifically bind and cleave mitochondrial DNA, which is subsequently degraded rapidly \citep{Peeva18}. Mathematical modelling of mitochondrially-targeted gene therapies predicts that tissues with high mean heteroplasmy and large heteroplasmy variance are generally more difficult to treat \citep{Hoitzing17}; there is therefore a close link between these promising therapeutic technologies and inter-cellular heterogeneity in heteroplasmy.

\section*{Non-genetic Sources of Mitochondrial Heterogeneity}

Apart from the mitochondrial genome, there are many non-genetic properties of mitochondria which can vary. One such property, which is particularly clear, is ETC protein structure which is influenced by sequence heterogeneity in mtDNA (Figure~\ref{Fig:cartoon}Bi) -- this being a potentially important source of both intra- and inter-cellular variability. Other non-genetic differences in mitochondria can include mitochondrial membrane composition/structure, ion content, membrane potential and network structure. Intra-cellular spatial heterogeneity through the existence of sub-cellular mitochondrial niches \citep{Palmer77,Mckenna00,Benador18}, e.g. perinuclear vs.\ peripheral locations, has been highlighted as an important axis of mitochondrial heterogeneity by previous authors, see \citep{Wikstrom09}. Below, we discuss several other aspects in which non-genetic attributes of mitochondria may vary, and the potential pathological consequences of such variation. We also draw attention to how non-genetic heterogeneity may be driven by, or drive, genetic heterogeneity through genotype-phenotype links. 

\subsection*{Stoichiometric and Structural Heterogeneity in the Inner Mitochondrial Membrane as a Potential Amplifier of the Genotype-phenotype Link}

\textit{MtDNA genotype as a driver of IMM phenotype}. The inner mitochondrial membrane (IMM) is heterogeneous in both its composition and its topology. The connection between the inner mitochondrial membrane and the mitochondrial genotype is particularly relevant since mitochondrial DNA is situated in the mitochondrial matrix and in close proximity to the IMM \citep{Brown11}. Consequently, one might expect mtDNAs to affect their local respiratory units more than distal mtDNAs \citep{Busch14}, thus allowing greater control of respiration at a local level \citep{Allen93,Allen03,Lane11}. Below we discuss aspects of how IMM physiology may generate mitochondrial heterogeneity, and how cardiolipin may act as a sensor of the mitochondrial genotype. We discuss how cardiolipin reacts to differences in reactive oxygen species (ROS) production and pH, and may amplify the mitochondrial phenotype through supercomplex formation and cristal structure.

\textit{Mitochondrial supercomplex stoichiometry as an axis of IMM heterogeneity}. The stoichiometry of respiratory units within the IMM can vary, since respiratory units organise into supramolecular structures termed supercomplexes \citep{Schagger00,Enriquez16} (Figure~\ref{Fig:cartoon}Biii). A variety of supercomplex stoichiometries exist in mammalian cells \citep{Schagger00,Schafer06} as well as free resipriatory subunits \citep{Schagger00}. The `plasticity model' has therefore been suggested \citep{Acin08}, whereby a landscape of combinations between respiratory complexes coexist. Dysfunction in the assembly of mitochondrial supercomplexes in mice results in decreased muscle activity and heat production in the cold due to reduced CIV activity \citep{Ikeda13}, showing that supercomplex assembly is required for fully-functional respiration. Consequently, dysfunction in supercomplex assembly could affect mitochondrial quality control pathways and influence the mitochondrial genotype.

\textit{Cristal structure as an axis of IMM heterogeneity}. The topology of mitochondrial membranes are heterogeneous \citep{Mannella06,Enriquez16}, and this heterogeneity may potentially be driven by mitochondrial genetic heterogeneity (see above), as we will discuss at the end of this subsection. ATP-synthase forms dimers in the inner mitochondrial membrane (IMM) which often arrange into micron-scale rows, which are associated with high local curvature to form mitochondrial cristae \citep{Strauss08,Davies12}. Recent work has suggested that the mitochondrial fusion protein OPA1 stabilises ATP synthase oligomers by modulating cristal shape \citep{Quintana18}. Crista membranes show an enrichment of respiratory complexes relative to the inner boundary membrane \citep{Gilkerson03,Vogel06}, and it has been suggested that cristae exist to increase the packing density of respiratory units \citep{Rieger14}. Individual cristae are morphologically heterogeneous in mammalian cells \citep{Frey00}, and can be modulated in response to altered metabolic demands \citep{Eisner18}: cristae become narrower in mammalian cells in response to starvation \citep{Patten14}, suggesting that cristal shape influences respiratory efficiency, perhaps by modulating local substrate concentrations \citep{Mannella01} (Figure~\ref{Fig:cartoon}Bii). Remodelling of mitochondrial morphology occurs during cell death \citep{Scorrano02,Yamaguchi08}, whereby supercomplexes and dimers of ATP synthase disassemble and cristal structure becomes disorganised \citep{Cogliati16}, allowing the release of cytochrome \textit{c} (an electron carrier of the ETC) to trigger the intrinsic cell death pathway \citep{Taylor08} (although the importance of crista remodelling for cytochrome \textit{c} release has been questioned \citep{Tam10}). The intimate connection between mitochondrial physiology and cell death provides insight into the recent observation that mitochondrial heterogeneity can partially explain variability in chemotherapeutic resistance in HeLa cells \citep{Marquez18}.

\textit{Cardiolipin is necessary for supercomplex and crista stabilisation}. Cardiolipin is a phospholipid found in the inner mitochondrial membrane, and stabilises both supercomplexes and cristal structure. When exposed to a pH gradient, cardiolipin-containing lipid vesicles spontaneously form crista-like membrane invaginations \citep{Khalifat08,Khalifat11}, thus providing a potential connection between respiratory activity, which is influenced by mtDNA genotype, and cristal shape. Furthermore, flies with deficient cardiolipin levels show reduced ATP synthase abundance in high-curvature regions of cristae, resulting in disorganised cristae, cardiac insufficiency, motor weakness and early death \citep{Acehan11}. In yeast, cardiolipin has been shown to be necessary for supercomplex stabilisation \citep{Zhang02,Pfeiffer03}. Indeed, patients with Barth syndrome who are deficient in cardiolipin due to a mutation in the tafazzin gene show both aberrant crista formation \citep{Acehan11} and reduced supercomplex formation \citep{Mckenzie06}. 

\textit{Cardiolipin as a potential amplifier of mtDNA heterogeneity through alterations in pH and ROS generation}. Cardiolipin is particularly susceptible to damage by ROS. In isolated bovine mitochondria, ROS exposure resulted in loss of CI activity, but exogenously added cardiolipin could restore CI activity \citep{Paradies02}. It has recently been shown that a mitochondrially-targeted antioxidant (MitoQ) is able to increase cardiolipin expression and content in liver mitochondria of rats fed on a high-fat diet, resulting in increased mitochondrial functionality and ATP synthase activity \citep{Fouret15}. Together, this suggests that mitochondrial dysfunction, which alters pH gradients across the IMM and ROS production, can cause cristae to become disorganised and affect supercomplex assembly via cardiolipin, potentially resulting in further loss of mitochondrial efficiency and ROS production. It is possible that heterogeneity in pH and ROS production, for instance through mtDNA mutation heterogeneity \citep{Lane11}, could be amplified through such mechanisms and thus strengthen the genotype-phenotype link between mtDNA and their local respiratory complexes. 

\subsection*{Mitochondrial Membrane Potential Heterogeneity, Mitochondrial Networks and Mitochondrial Genotype}

\textit{Inter-cellular mitochondrial membrane potential heterogeneity as a predictor of cell-physiological heterogeneity}. The inner membrane potential ($\dpsi$) is an indicator of mitochondrial functionality, generated by the ETC, which drives the synthesis of ATP by ATP-synthase. Mitochondrial output is highly sensitive to $\dpsi$: a 14~mV change in $\dpsi$ corresponds to a 10-fold change in the maximum ATP/ADP ratio \citep{Nicholls04}, where $\dpsi$ typically ranges between 150-180~mV \citep{Perry11}. Quantification of the absolute value of $\dpsi$ in millivolts at a single-cell level through fluorescence probes \citep{Perry11} is possible but technically challenging, requiring deconvolution from other confounding factors such as fluctuations in plasma membrane potential, the matrix:cell volume ratio, dye activity and binding affinity in the matrix/cytosol, and spectral changes resulting from binding \citep{Gerencser2012,Gerencser16}. Many studies involving $\dpsi$ measurements through Nernstian dyes neglect these possible confounding variables and assume differences in fluorescence are always directly attributable to differences in $\dpsi$, so some caution is required. With these caveats in mind, measurements of mitochondrial mass scaled by $\dpsi$ has been shown to explain much of the variation in transcript elongation rate \citep{Das2010,Johnston12} and protein noise \citep{Guantes15}, as well as predicting phenomena such as cell cycle duration \citep{Johnston12} and chemotherapeutic resistance \citep{Marquez18} in mammalian cells (Figure~\ref{Fig:cartoon}Bvi). 

\textit{Calcium and pH transients as determinants of intra-cellular $\dpsi$ fluctuations}. In addition to inter-cellular $\dpsi$ heterogeneity, individual mitochondria have been shown to undergo transient depolarisation/repolarisation cycles, termed `flickers' in animals \citep{Duchen98,Reilly03} or `pulses' in plants \citep{Schwarzlander12} (Figure~\ref{Fig:cartoon}Biv).  In freshly dissociated smooth muscle cells, flickers range from $<$10~mV to $>$100~mV, typically lasting on the order of seconds \citep{Reilly03}. It has been proposed that mitochondrial flickers are regulated by various mechanisms, including mitochondrial inner membrane fusion \citep{Santo13}, \ca{} influx \citep{Duchen98,Jacobson02}, and transient opening of the mitochondrial permeability transition pore (mPTP) \citep{Huser99,Jacobson02}. The mPTP is thought to be a non-selective mitochondrial channel that induces cell death when open for prolonged periods \citep{Bernardi06}. In rat myocytes, opening of the mPTP through pharmacological intervention has been found to correlate with the frequency of transients in circularly permuted yellow fluorescent protein (cpYFP) fluorescence \citep{Wang08}. cpYFP has been shown to be sensitive to pH \citep{Schwarzlander11,Santo13,Schwarzlander14}, therefore transient cpYFP fluorescence corresponds to transient alkalinisation of the mitochondrial matrix (i.e. an increase in pH gradient across the inner mitochondrial membrane). Since one would expect a reduction in pH gradient if the mPTP were behaving in a non-selective mode, transient alkalinisation potentially implies an ion-selective mode of the mPTP. The coincidence of an increase in pH gradient, and loss of membrane potential, indicates a redistribution of the proton motive force, which has pH and electrical contributions \citep{Nicholls04}. This redistribution may be mediated by selective ion movement. Flickers have been shown to exist \textit{in vivo} in mouse astrocytes, and are enhanced by both neuronal activity and oxidative shifts \citep{Agarwal17}, which is consistent with previous \textit{in vivo} observations in mice \citep{Breckwoldt14,Breckwoldt16}. Although no clear consensus exists on the mechanisms of transient mitochondrial depolarization/alkalinisation cycles, such cycles appear to be a likely means of regulating metabolic rate, and perhaps ROS production, at the single mitochondrion level.

\textit{Mitochondrial membrane potential influences mitochondrial genotype through quality control}. The canonical means by which mitochondrial membrane potential feeds back into the genetic state is through mitophagy and mitochondrial network dynamics \citep{Twig08}. Mitochondria are not static organelles but undergo dynamic fusion and fission, the purpose of which is incompletely understood \citep{Hoitzing15}. In rat pancreas cells, it has been shown that fission often results in a daughter mitochondrion which has a lower $\dpsi$ than its sister \citep{Twig08}. These depolarised mitochondria have a lowered propensity for fusion, and are more likely to be degraded \citep{Twig08}. Selective fusion, when combined with non-selective mitophagy, is sufficient to preferentially degrade depolarised/damaged mitochondria \citep{Hoitzing15,Aryaman18}, although the selective strength of mitophagy itself is not yet fully understood. The extent to which mitochondria which are degraded via selective/non-selective forms of mitophagy possess mitochondrial genomes which are perturbed from the consensus sequence also remains incompletely understood \citep{Lemasters14}. However, if depolarised mitochondria are more likely to be perturbed from the consensus mitochondrial sequence, mitochondrial quality control mechanisms would exert a negative selective pressure against variant alleles. Indeed, negative selection pressures have been observed in human tissues \citep{Li10,Avital12,Ye14,Li15} and at the intracellular level in mice \citep{Morris17}, which may be due to a combination of mitochondrial networks and mitophagy.

\textit{If quality control is weak, mitochondrial network fragmentation slows heteroplasmy variance through a rescaling of time}. The existence of mitochondrial diseases \citep{Schon12}, the ubiquity of heteroplasmy \citep{Payne12,Morris17}, and the accumulation of heteroplasmy with age \citep{Li15}, suggest that mitochondrial quality control may be weak for certain sequences. Building on insights from previous work \citep{Mouli09,Tam13,Tam15,Johnston16}, recent mathematical modelling \citep{Aryaman18} suggests that, if quality control is weak, heteroplasmy variance ($\mathbb{V}(h)$) follows the equation
\begin{equation}
\mathbb{V}(h) \approx f_s \frac{2 \mu t}{n} h_0(1-h_0) \label{eq:my_eq}
\end{equation}
where $t$ is time, $f_s$ is the fraction of unfused mitochondria, $\mu$ is the mitophagy rate, $n$ is copy number and $h_0$ is the initial heteroplasmy, which is equivalent to mean heteroplasmy under neutral drift (see Figure~\ref{Fig:dists_cartoon}). This equation arises through the assumption that larger mitochondrial fragments are at a reduced susceptibility to degradation, as is observed empirically \citep{Twig08}. As a consequence, total mitochondrial turnover is modulated by the fraction of unfused mitochondria, independently of the absolute magnitude of fusion-fission rates \citep{Aryaman18}. Since heteroplasmy variance is proportional to mitochondrial turnover through mitophagy \citep{Johnston16}, mitochondrial fragmentation may therefore modulate the rate of accumulation of pathologically mutated cells in a tissue, independently of selective effects \citep{Aryaman18} (Figure~\ref{Fig:cartoon}Bv). Mitochondrial network fragmentation may also slow de novo mutation \citep{Aryaman18} through a rescaling of mitochondrial turnover (which is known to modulate the de novo mutation rate \citep{Poovathingal09}). As a consequence, promoting mitochondrial fusion earlier in life, when mean heteroplasmy is low, may delay the rate of accumulation of cells with pathological levels of mutated mtDNA \citep{Aryaman18}, which may have implications for healthy ageing.

\subsection*{Mitochondrial Complementation May Dampen the Genotype-phenotype Link but its Extent is Incompletely Understood}

\textit{Mitochondrial complementation is thought to partially buffer genotype-phenotype links via the threshold effect}. Mitochondrial `complementation' consists of mitochondria sharing their contents through fusion-fission events, and potentially complementing each others' genetic defects \citep{Yoneda94,Hayashi94,Enriquez00,Ono01}. This is supported by experiments involving photoactivatable fluorescent proteins which show that intra-mitochondrial contents mix over time \citep{Twig08, Wilkens13}. Inhibition of fusion and fission have also been shown to induce heterogeneity in the distribution of aged mitochondrial proteins \citep{Ferree13}. Beyond depending on fission and fusion rates, and the degree of mitochondrial mobility, the strength of complementation is closely linked to the diffusivity of the mitochondrial matrix: if the matrix is a high-diffusivity environment then gene products are expected to be promiscuous within the matrix and not remain local to their parental mtDNA. Therefore, in cells heteroplasmic between wild-type and a pathological mutant, healthy versions of any particular transcript would be found in the matrix. The idea of complementation is also closely linked to the mitochondrial threshold effect \citep{Rossignol03,Stewart15,Aryaman17b}, whereby complementation effects allow cells to withstand high levels of mutant load (60-90\%, \citep{Chomyn92,Miyabayashi92,Rossignol03}) before displaying a respiratory defect.

\textit{Mitochondrial complementation remains incompletely understood}. Much of the progress in understanding mitochondrial complementation derives from cell fusion studies where cells harbouring different mtDNA mutations are fused, and mitochondrial functionality recovers through sharing of transcription products \citep{Ono01,Gilkerson08,Yang15}. Such experiments highlighted physiological subtleties which remain incompletely understood. For instance, the experiment of \cite{Ono01} between cells harbouring two different mt-tRNA mutations required an adaptive period of 10-14 days before respiratory activity was restored -- the reason for this was not fully understood. Furthermore, cell fusion studies between cells harbouring partially-functional (3271~T$>$C) and completely dysfunctional (3243~A$>$G) mitochondrial tRNA-Leu mutations resulted in a sigmoidal relationship between heteroplasmy and COX-activity, whereas a linear relationship was observed for simple co-culture between the two cell types \citep{Ono04}, which was not fully understood. In contrast to the experiment of \cite{Ono01}, an experiment by \cite{Gilkerson08} between two non-overlapping mitochondrial deletions showed recovery of MTCO2 protein production on a relatively fast timescale (4 days). Furthermore, \cite{Gilkerson08} found that fused cells cycled between heteroplasmic and homoplasmic states during long-term culture, despite being on a medium which mildly selected for mitochondrial function (but still allowed ATP production through glycolysis). Yet, a straightforward interpretation of complementation would suggest that heteroplasmic states would show maximal fitness. This effect was originally explained as a tension between a neutral genetic drift effect, deriving from the assumption that nucleoids possess several mtDNAs \citep{Jacobs00}, and selective pressure for heteroplasmy \citep{Gilkerson08}. However super-resolution microscopy studies have revealed that nucleoids harbour only 1-2 mtDNAs per nucleoid \citep{Kukat11,Kukat15}. It is possible that large drift rates could be explained by the passive partitioning of very large clusters of genetically homogeneous mtDNA at cell division. Furthermore, active asymmetric apportioning of damaged or aged mitochondria has been observed in yeast \citep{Mcfaline11} and stem cells \citep{Katajisto15} (which has been shown to influence cellular fate decisions in immune cells \citep{Adams16}); therefore, active mechanisms of asymmetric apportioning of mitochondrial genomes may also exist. Cycles between heteroplasmy and homoplasmy highlight an area for future experimental and theoretical investigation.

\textit{Restricted diffusion in the IMM may inhibit complementation}. Experiments by \cite{Wilkens12} fused HeLa cells with mitochondrially-targetted fluorescent proteins of differing colours to investigate the extent of mitochondrial diffusivity. The authors found that, through cycles of mitochondrial fusion and fission, IMM proteins appear to experience slow diffusion relative to the outer mitochondrial membrane, and retain cristal structure \citep{Wilkens13}. Consequently, complementation may be diminished since gene products may remain local to their parental mtDNA \citep{Busch14}, allowing local control of respiration \citep{Allen93, Allen03, Lane11}. This is supported by the observation that mitochondrial transcripts are particularly concentrated around mtDNA \citep{Ozawa07}. Mathematical modelling of progressive increases in heteroplasmy of the pathological 3243~A$>$G mutation is compatible with the interpretation that tRNAs are enriched in the vicinity of their local mtDNA, and that mutated mtDNAs experience a local depletion of ATP resulting in a transcriptional defect \citep{Picard14,Aryaman17b}. Hence, restricted complementation and local phenotype-genotype links appear to be explanatory.

\section*{Discussion}

We note that mitochondrial epigenetic modification is also a potential layer of non-genetic mitochondrial heterogeneity (Figure~\ref{Fig:cartoon}Bvii). Mitochondrial DNA can undergo methylation \citep{Van15b}; although its physiological impact is still being unravelled, it has been suggested that methylation may regulate mtDNA gene expression \citep{Van17}. Furthermore, whilst oxidative damage to mtDNA nucleotides have classically been known for their potential role in mutagenesis, it has been proposed that the formation of 8-OHdG, which is a ROS-modified version of guanine, induces mitochondrial mutations at the transcriptional level \citep{Nakanishi12} and may be responsible for premature ageing phenotypes in mice bred to rapidly accumulate mtDNA mutations \citep{Trifunovic04,Safdar11,Safdar16}. Post-transcriptional modifications of mitochondrial transcripts \citep{Bar13}, potentially modulated by heterogeneity in the sequence of nuclear DNA \citep{Hodgkinson14}, are also potential sources of mitochondrial heterogeneity, which potentially constitute a rich and fascinating avenue for future research. Further downstream sources of mitochondrial heterogeneity, such as translation errors, have also been shown to have physiological consequences in budding yeast \citep{Suhm18}. The age-related functional decline of mitochondria in shorter-lived animals \citep{Itsara14,Brandt17} might not be explained directly by mtDNA mutations \citep{Vermulst07,Lakshmanan18} (see also recent, contrasting, results in \textit{Drosophila} \citep{Kauppila18b,Samstag18}). However, the mechanisms described above which are downstream of mtDNA mutation, but are nevertheless constrained by mtDNA, may still be able to explain this functional decline.

Mitochondrial heterogeneity can occur at various scales. In this review, we have focussed on heterogeneity in the mitochondrial population within a cell (e.g.\ microheteroplasmy \citep{Morris17}) and inter-cellular heterogeneity of aggregate per-mitochondrion observables (e.g.\ variation in inter-cellular membrane potential \citep{Das2010,Johnston12}). Heterogeneity also exists at larger scales, for instance between organs of a particular individual (e.g.\ tissue-specific genetic selective pressures upon mtDNA \citep{Burgstaller14,Li15,Ahier18}), and heterogeneity between individuals (e.g.\ variation in the consensus sequence between individuals of different mitochondrial haplotypes \citep{Wallace13}). We have pointed out the genetic and non-genetic sources of such intra- and inter- cellular mitochondrial heterogeneity, as well as its pathological significance.

We have also discussed how genotype-phenotype links provide feedback between genetic and non-genetic states. The mitochondrial genotype may drive the phenotype through e.g. the genotype influencing ROS formation and pH gradients. This may, in turn, affect cardiolipin, supercomplex formation and cristal structure, which ultimately affect respiratory capacity. On the other hand, the mitochondrial phenotype may influence the genotype through e.g. the mitochondrial network and mitophagy. This may, in turn, modulate the mean and variance of the heteroplasmy distribution.

The interdependence between genetic and non-genetic sources of mitochondrial heterogeneity means that it is difficult to state which sources of mitochondrial heterogeneity are the most important in general. However, one may argue that genetic variability occurs at a slower timescale and may drive slowly-varying aspects of non-genetic mitochondrial heterogeneity. Slowly-varying aspects of mitochondrial heterogeneity may be especially important in explaining heterogeneous health outcomes during healthy ageing \citep{Lowsky13,Sun16,Kauppila17}. Therefore, understanding the dynamics of mitochondrial genetics through time has the potential to be particularly explanatory for age-related mitochondrial dysfunction. There exist various competing theoretical models which have the potential to describe mitochondrial genetic dynamics \citep{Chinnery99,Wonnapinij08,Poovathingal09,Johnston16,Aryaman18}; future interdisciplinary studies are required to constrain which of these models are statistically best able to describe experimental data (see e.g. \cite{Johnston15} for an example of how statistical inference can constrain different theoretical explanations of the mitochondrial bottleneck).

In order to deepen our understanding of mitochondrial genetics, it is important for experimental studies to shift away from heteroplasmy measurement in bulk cellular samples, and towards single-cell studies. In bulk measurements, there is no way to determine whether individual cells are homoplasmic or heteroplasmic: this has important consequences for development and mitochondrial inheritance \citep{Johnston15,Johnston15b}. Furthermore, in the context of variations in mtDNA copy number between experimental conditions, inferences about differences in heteroplasmy in tissue homogenate can be erroneous and apparently display selective effects where there are in fact none \citep{Hoitzing17}. Such difficulties may be circumvented when heteroplasmy is measured at the single-cell level.

We have highlighted several other important outstanding questions in the field of mitochondrial heterogeneity:
\begin{itemize}
\item What explains the apparent ubiquity of mtDNA copy number density homeostasis, despite mtDNA density being a potential axis of energetic control?
\item What is the extent, and physiological importance, of microheteroplasmy (Fig.~\ref{Fig:cartoon}Aii)?
\item To what extent do inherited mtDNA mutations, versus somatic mtDNA mutations, contribute to healthy ageing?
\item To what extent does mitochondrial exchange affect heteroplasmy dynamics?
\item To what extent can non-genetic mitochondrial heterogeneity in factors such as pH, ROS production, and cristae structure be explained by genetic mitochondrial heterogeneity?
\item Under what circumstances, and to what extent, is mitophagy selective under physiological conditions?
\item To what extent are mtDNAs asymmetrically partitioned amongst proliferating cells \textit{in vivo}?
\item To what extent may mtDNAs complement each others genetic defects through sharing of gene products?
\end{itemize}   
\noindent{}The richness in sources of mitochondrial heterogeneity, as well as the growing appreciation of its pathophysiological importance, will likely provide future insight into the determinants of cellular heterogeneity and its associated pathologies.

\section*{Conflict of Interest Statement}

The authors declare no competing interests.

\section*{Author Contributions}

NSJ conceived the project. JA performed the literature review and wrote the manuscript with input from IGJ and NSJ.

\section*{Funding}

JA acknowledges grant support from the BBSRC (BB/J014575/1) and the MRC Mitochondrial Biology Unit (MC\textunderscore{}UP\textunderscore{}1501/2). IGJ acknowledges support from the University of Birmingham via a Birmingham Fellowship. NSJ acknowledges grant support from the BHF (RE/13/2/30182) and EPSRC (EP/N014529/1).

\section*{Acknowledgments}
We would like to thank Markus Schwarzl\"{a}nder, Hanne Hoitzing, Benjamin Ingledow and Ferdinando Insalata for useful discussions, and the contributors to the \url{imperialmitochondriacs.blogspot.com} blog.

%%% If you are submitting a figure with subfigures please combine these into one image file with part labels integrated.
%%% If you don't add the figures in the LaTeX files, please upload them when submitting the article.
%%% Frontiers will add the figures at the end of the provisional pdf automatically
%%% The use of LaTeX coding to draw Diagrams/Figures/Structures should be avoided. They should be external callouts including graphics.

\end{document}